\documentclass[10pt,a4paper,aps,twocolumn,showpacs,superscriptaddress,nofootinbib]{revtex4-1}
\usepackage[colorlinks,hyperindex]{hyperref}
\usepackage{epsfig}
\usepackage{float}
\usepackage{float,upgreek,yfonts}
\RequirePackage{amssymb,amsmath}
\usepackage[latin1]{inputenc}
\usepackage{graphicx}
\usepackage{indentfirst}
\usepackage{color}
\begin{document}
\title{Nuclear configurational entropy of the energy-energy correlation in $e^+e^-$ annihilation processes}
\author{G. Karapetyan}
\email{gayane.karapetyan@ufabc.edu.br}
\affiliation{Centro de Ci\^encias Naturais e Humanas, Universidade Federal do ABC - UFABC, 09210-580, Santo Andr\'e, Brazil}
\begin{abstract}
In this paper the nuclear configurational entropy of the energy--energy correlation function is scrutinized, in $e^+e^-$ collisions to hadrons in
the back-to-back region,  within the QCD perturbative theory.
The critical points of the nuclear configurational entropy concept,  in  the hot nuclear system, is shown to correspond to the critical angle between the two scattered hadrons, corroborating with phenomenological data with great accuracy. The main tools employ  the main coefficients of the soft-gluon resummation formula, taking into account logarithmically enhanced contributions up to next-to-next-to-leading logarithmic accuracy.
\end{abstract}
\maketitle
\section{Introduction}

The mass generation in high energy hadron reactions within Quantum Chromodynamics (QCD) has been spectacularly
confirmed by experimental data and remains an important task of particle physics.
However, there are several questions regarding the mechanism of the reactions, that are required to be understood.
Particularly, one of the interesting topics is the QCD predictions concerning $e^+e^-$ annihilation into hadrons
which allows to obtain the energy-energy correlation as well as  to measure the strong coupling constant.
Using the configurational entropy (CE) techniques gives an opportunity to test the new experimental data of QCD, as well as to corroborate with the existing ones. It also serves  as one of the most precise tools to predict and to find appropriate parameters to describe the different phenomena of high energy nuclear systems.
The study of the CE critical points in the frame of dynamical AdS/QCD models allowed to explain and quantitatively obtain the light-flavour mesons production with lower spin value \cite{Bernardini:2016hvx}.
Another good experience is connected with the prevalence of the lower-spin bound states of gluons, the glueballs, as a consequence of the self-coupling of gluons in QCD \cite{Bernardini:2016qit}. The CE in the QCD context has been recently used to derive information-entropic Regge trajectories for meson families in the seminal Ref. \cite{Bernardini:2018uuy}, whereas the dissociation of heavy mesons, bottomonium  and charmonium, was scrutinized in Refs. \cite{Braga:2017fsb,Braga:2018fyc}.
There are also many aspects when the CE critical points can provide the useful information about the relative stability of the physical configurations, as in  Bose--Einstein  condensates \cite{Casadio:2016aum}.  Using of the conceptual setup of the CE permitted to deeper understand and explain such a phenomena, as the Hawking--Page transition \cite{Braga:2016wzx}, as well as to obtain the KdV in a  quark-gluon plasma \cite{daSilva:2017jay}.  Light-flavour mesonic states with tachyonic corrections were studied in Ref. \cite{Barbosa:2018mng}.

We would like to stress out that one can use CE approach
in order to predict the critical points of any localized nuclear configuration  \cite{Colangelo:2018mrt}. It is based on the problem of finding the critical points of CE, as a convenient tool for investigating
the configurational stability which takes into account the dominant role of the physical systems under study \cite{Gleiser:2018kbq,Gleiser:2012tu, Sowinski:2015cfa, Gleiser:2014ipa, Gleiser:2011di, Gleiser:2015rwa,Gleiser:2013mga}, as it was recently done for the analysis of the complex nuclear configurations in order to describe the character of the strong interaction in the frame of QCD \cite{Ma:2018wtw,Ma:2015lpa,Bazeia:2018uyg}.
Since the reaction cross section represents the probability that a nuclear reaction occurs, being a  spatially localized function,
it is possible calculate the Fourier transform of the  reaction cross section and thus obtain  the nuclear configurational entropy (NCE) \cite{Karapetyan:2018yhm,Karapetyan:plb,Karapetyan:2017edu,Karapetyan:2016fai}. Its critical points corroborated
with a wide range of experimental data in QCD. In addition, aspects of the CE were also used in particle physics, in Ref.
\cite{Alves:2014ksa}.

The NCE paradigm can serve as a great tool in QCD theory, which studies the nuclear interactions among partons at high energy range, showing a good agreement of experimental data and theory.
Measuring the critical points in the functional space which comprise all energy of the nuclear configurations, the NCE brings all the information concerning the reaction mechanism.
 The NCE also can be then employed to find the energy--energy correlation function in $e^+e^-$ annihilation into hadrons in back-to-back region \cite{Florian}.
In above mentioned region, the resummation of the logarithmic contributions to the region of small transverse momenta in the distributions of high-mass systems produced in hadron collisions is considered.
In the perturbative QCD, the resummation formulae that are usually used to calculate the distributions, includes process-dependent form factors and coefficient functions.
The knowledge of the coefficients leads to the resummation of the all leading logarithmic contributions, and
allows us to perform an excellent matching of the resummed and fixed--order calculations.

The outline of our paper is as follows: in the next section the energy--energy correlation function in $e^+e^-$ collisions to hadrons is introduced.
We briefly discuss the general considerations
of fixed--order calculations.
In section III we review the resummation formalism
representing the Soft-gluon resummation formula together with the main coefficients and form factors used in the calculation, as well as the outline of the formalism to determine the critical point of the configurational entropy approach.
We briefly summarize our results in a
concluding section.

\section{Energy--energy
correlation in $e^+e^-$ collisions}

At the QCD high energy regime,
 observables as the energy-energy correlation (EEC) and the
strong coupling constant $\alpha_s$ have been 	
tellingly obtained due to precise experimental data concerning
$e^+e^-$ annihilation \cite{Florian}.
In this setup, the annihilation reaction is based upon the channel
$e^++e^-\to h_a+h_b +X$. It can be quantitatively characterized by
the energy-weighted correlation for the cross section, EEC, reading
\begin{equation}
\label{eecdef}
\frac {1}{\upsigma_{\rm T}}\frac{d\Upsigma}{d\cos\upchi} = \frac{1}{\upsigma_{\rm T}}
\sum_{a,b}\int \frac {E_a\, E_b}{Q^2}
\updelta(\cos\upchi+\cos\uptheta_{ab})\,d \upsigma,
\end{equation}
where $d \upsigma$ denotes the cross section of the above mentioned channel.
In Eq.~(\ref{eecdef})
$\upsigma_{\rm T}$ is the total hadronic cross section,
$Q$ represents the energy of the centre-of-mass,
$\uptheta_{ab}=\pi-\upchi$
is the angle between the particles $a$ and $b$,
and finally $E_a$ and  $E_b$ are the energies of the particles in the exit reaction channel.
For the back-to-back region, hence the angle $\uptheta_{ab} \to \pi$  corresponds to $\upchi \to 0$, whereas
the normalization condition provides the EEC distribution integral in Eq.~(\ref{eecdef}) must be unity at $0\leq \upchi \leq \pi$.

The multiple collinear gluons emission in the back-to-back region lead  to enhancing the logarithmic contribution, which can be expressed as $\alpha_s^n \log^{2n-1}y$ with $y=\sin^2\upchi/2$.
The logarithmic term tends to increase as $y$ decreases, and to utilize the fixed-order perturbative expansion becomes unavailable.
The state of the art currently includes fixed-order next-to-next-to-leading order (NNLO) correction. In this context, the logarithmic contributions can be resummed over all orders.
The resummation procedure implies an analogy between the formalism for the transverse-momentum distribution of high-mass systems in hadronic collisions, when the transverse momentum can be replaced by the $q_{\rm T}^2=Q^2\sin^2(\upchi/2)$ observable for EEC calculation.
This procedure implies also that the large logarithmic contributions $\alpha_s^n \log^{2n-1} q_{\rm T}^2/Q^2$ can be resummed to all orders, in the case when the final hadronic state invariant mass $Q^2$ is higher than its transverse momentum $q_{\rm T}^2$.
The computation of the NNLO correction for EEC in $e^+e^-$ annihilation reaction
leads to the calculation of the coefficients that are responsible for the given order in the case of the presence the same order analytical estimation by fixing still absent coefficients at
${\cal O}(\alpha_s^2)$ (logarithmically enhanced contributions) and including the next-to-next-to-leading logarithmic (NNLL)
precision to resummation procedure for the back-to-back region.

\section{The resummation formalism}

According to Ref. \cite{Florian}, the energy-energy correlation can be represented by the following expression:
\begin{equation}
\label{deco}
\frac {1}{\upsigma_{\rm T}}\frac {d \Upsigma}{d\cos\upchi}=
\frac {1}{\upsigma_{\rm T}}\frac {d \Upsigma_1}{d\cos\upchi}
+\frac {1}{\upsigma_{\rm T}}\frac {d \Upsigma_2}{d\cos\upchi}\, ,
\end{equation}
where the first term includes
all logarithmically enhanced contributions, $\alpha_s^n/y \log^m y$. To calculate it, one should resum them to all orders.
The second term has no above mentioned contribution and the fixed-order perturbation theory can give the quantitatively estimation of it.

The first term of the Eq.~(\ref{deco}) then can be represented using
Sudakov form factor as:
\begin{equation}
\label{CS}
\frac{1}{\upsigma_{\rm T}}\frac{d \Upsigma_1}{d\cos\upchi}=
 H(\alpha_s(Q^2))\frac{Q^2}{8}
\int_0^\infty b\, J_0(bq_{\rm T})
S(Q,b)\, db\,  ,
\end{equation}
where $b$ is the impact parameter, $H(\alpha_s(Q^2))$ is the function that includes
hard contributions from virtual corrections at scale $q\sim Q$,
$J_0(bq_{\rm T})$ is the Bessel function, and $S(Q,b)$ is the form factor, which takes into account the
virtual and real contributions from
soft (the function $A=A(\alpha_s(q^2))$ and
flavour-inclusive
collinear (the function $B=B(\alpha_s(q^2))$ radiation at the scales $1/b\leq q \leq Q$ in the following form \cite{Florian1}:
\begin{equation}
\label{sudakov}
S(Q,b)\!=\!\exp \left\{ \!-\!\int_{\frac{b_0^2}{b^2}}^{Q^2} \frac{dq^2}{q^2}\!
\left[ A\ln \frac{Q^2}{q^2} \!+\! B \right] \right\}\, .
\end{equation}
In  Eq.~(\ref{sudakov})
the parameter $b_0=2 e^{-\gamma_E}$
has a kinematical origin.
Eq.~(\ref{CS}) indicates that in the case of back-to-back localization of the triggered partons, the radiation that is emitted during the reaction  is suppressed and thus in this case just soft and collinear partons are released.

One can use a perturbative
expansions in $\alpha_s$ in order to calculate the functions $A$, $B$, $H$ (Eqs.~(\ref{CS}), (\ref{sudakov})), which  are free of logarithmic corrections:
\begin{eqnarray}
\label{aexp}
A(\alpha_s) &=& \sum_{n=1}^\infty \left( \frac{\alpha_s}{\pi} \right)^n A^{[n]} \;\;, \\
\label{bexp}
B(\alpha_s) &= &\sum_{n=1}^\infty \left( \frac{\alpha_s}{\pi} \right)^n B^{[n]}
\;\;, \\
\label{cexp}
H(\alpha_s) &=& 1 +
\sum_{n=1}^\infty \left( \frac{\alpha_s}{\pi} \right)^n H^{[n]} \;\;.
\end{eqnarray}

After the integration procedure of Eq.~(\ref{sudakov}) over $q^2$,
one can obtain the form factor as:
\begin{align}\label{sudd}
S(Q,b)=\exp\left(L \, g_1 + g_2 + a_{\mathrm{S}}\, g_3+\cdots\right)\, ,
\end{align}
where $g_i=g_i(\alpha_{\rm S}\upbeta_0L)$, $a_{\mathrm{S}}=\alpha_s/\pi$
and $L=\log Q^2 b^2/b_0^2$ is the large logarithm, which is equal to $\log y$ at large values of the impact parameter $b$.

The functions $g_i(\upxi)$, where $\upxi$ is a saturation exponent, can be represented as \cite{Florian}:
\begin{widetext}
\begin{align}
\label{gfunctions}
g_1(\upxi) &= \frac {A^{[1]}}{\upbeta_0} \left(
\frac{\log(1-\upxi)}{\upxi} +1\right)\nonumber \\
g_2(\upxi) &= \frac {B^{[1]}}{\upbeta_0} \log(1-\upxi) -\frac {A^{[2]}}{\upbeta_0^2} \left(\log(1-\upxi)+ \frac {\upxi}{1-\upxi}\right) + \frac {A^{[1]}}{\upbeta_0} \left( \log(1-\upxi) +\frac {\upxi}{1-\upxi}\right) \log\frac {Q^2}{\mu_R^2}  \nonumber \\
& +\frac {A^{[1]} \upbeta_1}{\upbeta_0^3} \left( \frac {\upxi}{1-\upxi}+\frac{\log(1-\upxi)}{1-\upxi} +\frac {1}{2} \log^2(1-\upxi)\right) \nonumber \\
g_3(\upxi) &= -\frac {A^{[3]}}{2 \upbeta_0^2} \frac {\upxi^2}{(1-\upxi)^2}
-\frac {B^{[2]}}{\upbeta_0} \frac {\upxi}{1-\upxi}
+\frac {A^{[2]} \upbeta_1}{\upbeta_0^3(1-\upxi)^2} \left( \frac {\upxi}{2} (3\upxi-2)
 - (1-2\upxi) \log(1-\upxi) \right) \nonumber \\
& + \frac {B^{[1]} \upbeta_1}{\upbeta_0^2} \left(\frac {\log(1-\upxi)}{1-\upxi}+\frac {\upxi}{1-\upxi} \right) - \frac {A^{[1]}}{2} \frac {\upxi^2}{(1-\upxi)^2}  \log^2\frac {Q^2}{\mu_R^2} \nonumber \\
&+ \log\frac {Q^2}{\mu_R^2} \left(A^{[1]} \frac {\upbeta_1}{\upbeta_0^2}
 \left(\frac {1-2\upxi}{(1-\upxi)^2} \log(1-\upxi)+\frac {\upxi}{1-\upxi}+B^{[1]} \frac {\upxi}{1-\upxi} + \frac {A^{[2]}}{\upbeta_0}
 \frac {\upxi^2}{(1-\upxi)^2} \right) \right) \nonumber \\
&  +A^{[1]} \left( \log(1-\upxi)+ \frac {\upbeta_1^2}{2 \upbeta_0^4} \frac {1-2\upxi}{(1-\upxi)^2} \log^2(1-\upxi)
\left[  \frac {\upbeta_0 \upbeta_2 -\upbeta_1^2}{\upbeta_0^4} +\frac {\upbeta_1^2}{\upbeta_0^4 (1-\upxi)}  \right] \right.  \nonumber \\
& \left. + \frac {\upxi}{2 \upbeta_0^4 (1-\upxi)^2} ( \upbeta_0 \upbeta_2 (2-3\upxi)+\upbeta_1^2 \upxi)
 \right)
\end{align}
where $\upbeta(\alpha_s)$ is the QCD $\upbeta$-function:
\begin{align}
\upbeta_0 &= \frac{1}{12} \left( 11 c_A - 2 n_f \right) \;\;,
\quad\quad \upbeta_1=  \frac{1}{24}
\left( 17 c_A^2 - 5 c_A n_f - 3 c_F n_f \right) \;\;,
\nonumber \\
\label{bcoef}
\upbeta_2 &= \frac{1}{64} \left( \frac {2857}{54} c_A^3
- \frac {1415}{54} c_A^2 n_f - \frac {205}{18} c_A c_F n_f + c_F^2 n_f
+ \frac {79}{54} c_A n_f^2 + \frac {11}{9} c_F n_f^2 \right) \;\;.
\end{align}
\end{widetext}

During the EEC function calculation, the $g_1$, $g_2$, $g_3$ functions responsible for
the leading logarithmic (LL), next-to-leading logarithmic (NLL) and next-to-next-to-leading logarithmic (NNLL) accuracy, correspondingly.
Thus, the knowledge of coefficients $A^{[1]}$ leads to the resummation of the LL contributions. Analogously, the coefficients $A^{[1]}$, $A^{[2]}$ and $B^{[1]}$ give the NLL terms \cite{Kodaira}, the coefficients $A^{[1]}$, $A^{[2]}$, $A^{[3]}$, $B^{[1]}$, $B^{[2]}$ give the NNLL terms, and so forth, where the coefficient
$A^{[3]}$, the leading soft term in the three-loop splitting functions, has been taken from Ref.\cite{Vogt}.
They have the same form as in the quark form factor in the transverse
momentum distributions for hadron-hadron interaction:
\begin{align}
\label{coef1}
A^{[1]}&= c_F \nonumber \\
B^{[1]}&= -\frac {3}{2} c_F \nonumber \\
A^{[2]}&=
\frac {1}{2}\Big(c_A\,\left(\frac {67}{18}-\frac {{\pi }^2}{6}\right)-\frac {5}{9}n_f \Big)\,\, A^{[1]} \, .
\end{align}
We can utilize next-to-leading order (NLO) equation for the total reaction cross section, $\upsigma_{\rm T}$:
\begin{equation}
\upsigma_{\rm T}^{\rm NLO}= \frac {4\pi \alpha^2}{Q^2} \sum_q e_q^2  \left(1+\frac {\alpha_s}{2\pi}\, \frac {3}{2}\, c_F\right)\, ,
\end{equation}
and the coefficient $H^{[1]}$:
\begin{equation}
\label{H1}
H^{[1]}=  -c_F \left(\frac {11}{4}+\frac {\pi^2}{6}\right) \,.
\end{equation}
For the $A^{[3]}$ coefficient, one can use the expression:
\begin{align}
A^{[3]}=\frac {1}{4}&\left[ c_A^2 \left( \frac {245}{24} -\frac {67}{9} \zeta_2 +\frac {11}{6} \zeta_3 +\frac {11}{5} \zeta_2^2 \right)  \right. \nonumber\\
& \left.+c_F  n_f \left( -\frac {55}{24} +2 \zeta_3 \right) \right. \nonumber\\
& \left. +c_A n_f \left( -\frac {209}{108} +\frac {10}{9} \zeta_2 -\frac {7}{3} \zeta_3 \right)
- \frac {1}{27} n_f^2
\right] \, A^{[1]}\, ,
\end{align}
with $\zeta_n$ being the Riemann $\zeta$ function.
The resummation coefficient $B^{[2]}$ for NNLL accuracy can be obtained by the comparison of the
the logarithmic structure of the EEC fixed-order
perturbative calculation and
the expansion of the resummed component for the EEC function  Eq.~(\ref{CS}):
\begin{widetext}
\begin{align}
\label{expansion}
\frac {1}{\upsigma_{\rm T}} \frac {d\Upsigma^{\rm (res.)}}{d\cos\upchi}=&
\frac {1}{4y}\Bigg\{\frac {\alpha_s}{\pi}
\Big[ -{A}^{[1]} \log y +{B}^{[1]} \Big]
+ \left(\frac {\alpha_s}{\pi}\right)^2
\Bigg[
 \frac {1}{2} \left( {A}^{[1]}\right)^2 \log^3 y
+\left(-\frac {3}{2} {B}^{[1]}  {A}^{[1]} +
\upbeta_0 {A}^{[1]} \right) \log^2 y  \nonumber\\
&+ \left( -{A}^{[2]} -
\upbeta_0 {B}^{[1]}
 +  \left(  {B}^{[1]} \right)^2
 - {A}^{[1]} {H}^{[1]}  \right) \log y  + {B}^{[2]}
 +  {B}^{[1]} {H}^{[1]}
  +2 \zeta_3 ({A}^{[1]})^2 \Bigg] +{\cal O}(\alpha_s^3) \Bigg\}\, ,
\end{align}
\end{widetext}
with $\mu_R=Q$.

In order to obtain the result for the EEC, it is enough to calculate its small-$y$ behaviour, which obeys the same procedure as for the transverse-momentum distribution in hadronic collisions.
In order to obtain the universal functions for the infrared structure of elements of the QCD matrix at ${\cal O}(\alpha_s)$, one can use improved factorization formulae \cite{Florian} to compute the small-$y$  behaviour of EEC in a simpler manner.
For the leading-order subprocess with the production of a $q{\bar q}$ pair
the coefficient $B^{[2]}$ is given by:
\begin{equation}
\label{B2eecf}
B^{[2]}= -\frac {1}{2} \gamma_{q}^{[2]} + c_F \upbeta_0\left(\frac {5}{6}\pi^2-2\right)\, ,
\end{equation}
with
the coefficient $\gamma_q^{[2]}$ being
\begin{eqnarray}
\gamma_{q}^{[2]}&=& c_F^2\left( \frac {3}{8}-\frac {\pi^2}{2}+6 \zeta_3 \right) +
 c_F c_A \left( \frac {17}{24}+ \frac {11 \pi^2}{18}-3 \zeta_3\right)\nonumber\\&&
- c_F n_f T_R\left( \frac {1}{6}+ \frac {2 \pi^2}{9}\right) \, .
\end{eqnarray}

Now, in order to obtain the critical points of NCE,
one starts from the computing the spatial Fourier transform of the energy--energy correlation function cross section \eqref{eecdef}, to derive and compute the modal fraction.
As we already mentioned, the CE techniques involves the concept of the spatially-localised functions,
\cite{Gleiser:2012tu,Gleiser:2014ipa,Gleiser:2011di,Gleiser:2013mga}, as here the EEC.
In high energy correlated systems of hadrons, the main observable is the reaction cross section that is also
also spatially-localised and describe the entire reaction mechanism \cite{Karapetyan:2018yhm,Karapetyan:plb,Karapetyan:2017edu,Karapetyan:2016fai}. Hence one can take the EEC and compute the associated NCE.

The first thing that should be calculated is the Fourier transform of the energy--weighted correlation for the cross section corresponding to the $e^++e^-\to h_a+h_b +X$ process, into the $k$ momenta space:
\begin{equation}
\label{34}
\frac {1}{\upsigma_{\rm T}}\frac{d\Upsigma(\upchi,k)}{d\cos\upchi}{}=\frac{1}{2\pi}\! \int_{-\infty}^\infty \frac {1}{\upsigma_{\rm T}}\frac{d\Upsigma(\upchi,b)}{d\cos\upchi}\, e^{ikb} d b\,\,.
\end{equation}  The respective modal fraction is then computed as,
\begin{equation}\label{modall}
f_{\Upsigma(\upchi,k)}=\frac{\Big\vert\,\frac {1}{\upsigma_{\rm T}}\frac{d\Upsigma(\upchi,k)}{d\cos\upchi}{}\,\Big\vert^2}{\int_{-\infty}^\infty\Big\vert\,{\frac {1}{\upsigma_{\rm T}}\frac{d\Upsigma(\upchi,k)}{d\cos\upchi}{(k)}\Big\vert^2}\,dk}.
\end{equation}
It represents the amount of information of the system encoded into
its $k$ mode. It also measures the contribution of any given $k$ mode to the so called power spectrum \cite{Bernardini:2018uuy}.
The NCE can be thus derived as \begin{equation}
\label{333}
S_\Upsigma(\upchi) \,= \, - \int_{-\infty}^\infty f_{\Upsigma(\upchi,k)} \log  f_{\Upsigma(\upchi,k)} d k.
\end{equation}
Eqs. (\ref{34}) - (\ref{333}) for the NCE can be computed for the energy--energy correlation function in the
$e^+e^-$ annihilation process \eqref{eecdef}.

The quantitative results of the EEC (\ref{eecdef}) as a function of the $\upchi$ angle, at NLL+LO accuracy obtained using Eqs. (\ref{CS}), (\ref{sudakov}) and (\ref{sudd}) by fixing $\Lambda^{n_F=5}_{\rm QCD}=0.1665$ GeV, corresponding to $\alpha_s(M_Z)=0.130$ and $\mu_R=Q$ were shown in Fig. 1 of Ref. \cite{Florian} with a  solid line. There is a maximum critical point of the EEC at $\chi \approx 4.3^{\circ}$, where the EEC, given by Eq. (\ref{eecdef}), equals to 1.63. This  peak value then drops steeply, up to asymptotic value
$\upchi \geq 60^{\circ}$.
It can be noted  that the matching term  becomes significant up to very small values of $\upchi$ and is turned to be dominant at larger $\upchi$.

\begin{figure}[t]
\includegraphics[scale=0.6]{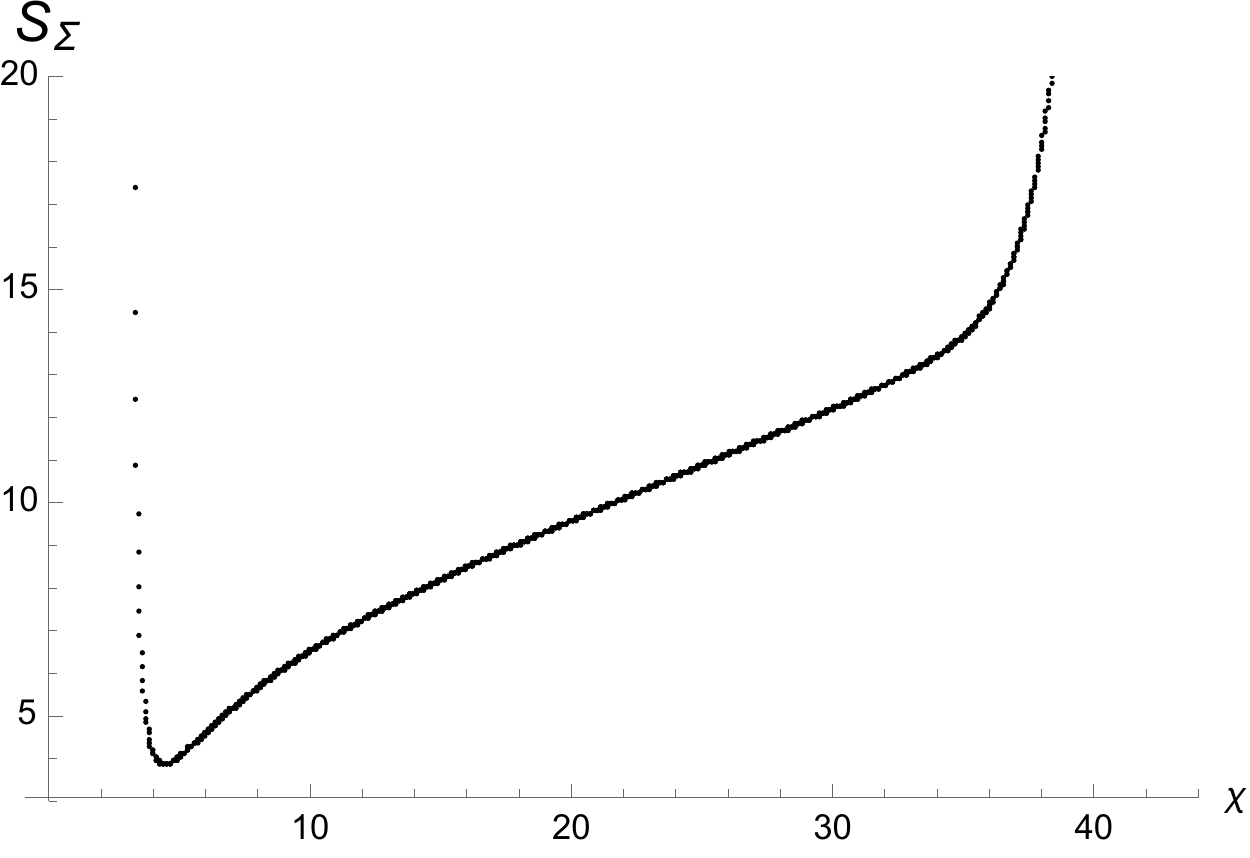}
\caption{Plot of the nuclear configurational entropy (NCE) (\ref{333}) of the energy-energy correlation (EEC), $\frac {1}{\upsigma_{\rm T}}\frac{d\Upsigma}{d\cos\upchi}$, as a function of the angle $\upchi$, defining the angle between the two scattered hadrons.
}\label{fig1}
\end{figure}

The results obtained for the nuclear CE, for the for the energy--energy correlation function, show an excellent agreement with phenomenological data, for the coefficients in a leading-order subprocess \cite{Florian}.
Computed by Eq. (\ref{333}), with $\Lambda^{n_F=5}_{\rm QCD}=0.1665$ GeV, $\alpha_s(M_Z)=0.130$ and $\mu_R=Q$, the NCE is then plot in Fig. \ref{fig1}.
It is worth to emphasize that the minimum of the NCE is given by $\chi \approx 4.22^{\circ}$, with 1.7\% of accuracy with phenomenological data \cite{Florian}. In addition, Fig. \ref{fig1} physically corroborates
with the solid line of Fig. 1 in Ref. \cite{Florian}, whose EEC attains low values for $\chi\to0$ and $\chi\geq 60^{\circ}$. In fact, the critical points of the NCE $S_\Upsigma(\upchi)$ refer to the
the configurations of the hot nuclear system that correspond to a higher organization of its modes, wherein the information is more compressed. The minimal point ($\chi \approx 4.22^{\circ}; S_\Upsigma \approx 2.42$) in Fig. \ref{fig1} also
extremizes the EEC from an informational perspective.
Hence it corresponds to the most probable array of the nuclear system configuration. It is in full agreement with the solid line in Fig. 1 of \cite{Florian}, wherein a maximum critical point of the EEC is at $\chi \approx 4.3^{\circ}$

Using the concept of the Shannon's information entropy one can figure out the critical point of the nuclear CE and thus establish the natural occurrence of any observable, here the NCE of the EEC.
It was determined by the critical point of the NCE, providing
a platform in nuclear systems for predicting and corroborating the  value of the EEC as a function of any system parameter. Here
the chosen parameter was the $\upchi$ angle, that defines the angle of the scattered hadrons in $e^+e^-$ channels. The derived critical point of the NCE then corresponds to the dominant states of the system nuclear configurations.

\section{Conclusions}

In the framework of QCD perturbative theory, the coefficients in the leading-order subprocesses constituted an useful tool for computing the NCE of the energy--energy correlation. As the EEC is a localized, square integrable scalar field, measuring the energy-weighted correlation for the cross section, the NCE paradigm developed in Refs. \cite{Karapetyan:2018yhm,Karapetyan:plb,Karapetyan:2017edu,Karapetyan:2016fai} could be forthwith emulated,  
for the energy-energy correlation in the $e^+e^-$ annihilation processes. The computed NCE $S_\Upsigma$ as a function of the $\upchi$ angle,  in Fig. \ref{fig1}, has a global minimum
$S_\Upsigma \approx 2.42$, corresponding to the angle $\chi \approx 4.22^{\circ}$. Eqs. (\ref{34}) - (\ref{333}) were applied to calculate the critical points of the CE. Our results corroborates within 1.7\% of accuracy with the phenomenological data in Ref. \cite{Florian}. The asymptotic values of the $\upchi$ angle also physically endorse the phenomenological results in Fig. 1 in Ref. \cite{Florian}. Since this global minimum of  the NCE $S_\Upsigma(\upchi)$ corresponds to the
 the system configurations whose underlying information is more compressed, the value $\chi \approx 4.22^{\circ}$ yields a maximum 
 of the energy-weighted correlation for the cross section, $\frac {1}{\upsigma_{\rm T}}\frac{d\Upsigma}{d\cos\upchi}$, as illustrated in  
 Fig. 1 of \cite{Florian}, wherein a maximum critical point of the EEC is at $\chi \approx 4.3^{\circ}$.

Using the concept of the Shannon's information entropy one can figure out the critical point of the nuclear CE and thus establish the natural occurrence of any observable, here the NCE of the EEC.
It was determined by the critical point of the NCE, providing
a platform in nuclear systems for predicting and corroborating the  value of the EEC as a function of any system parameter. Here
the chosen parameter was the $\upchi$ angle, that defines the angle of the scattered hadrons in $e^+e^-$ channels. The derived critical point of the NCE then corresponds to the dominant states of the system nuclear configurations.
During the calculation procedure the minima of the configurational entropy could predict the predominant nuclear states, providing the natural set of the observables and show an excellent agreement with  theoretical and phenomenological data. 
We intend to continue the study of nuclear configurations by involving different  QFT effects in the context of the AdS/QCD setup, using the dilatons and warp factors in Refs. \cite{Bazeia:2013usa,Bazeia:2012qh,Correa:2016pgr,Correa:2015vka,German:2012rv,German:2013sk}, in the NCE setup.

  \acknowledgements
GK thanks to FAPESP (grant No. 2016/18902-9), for partial financial support.

\end{document}